\documentclass[twocolumn]{webofc}
\usepackage[varg]{txfonts}   
\newcommand{\Xmax}{X_{\rm max}}

\newcommand{\dEdX}{\mathrm{d}E/\mathrm{d}X}
\newcommand{\dEdXmax}{\left(\dEdX\right)_{\rm max}}

\begin{document}
\title{Average shape of longitudinal shower profiles measured at the Pierre Auger Observatory}
%
%

\author{
\firstname{Sofia} \lastname{Andringa}\inst{1}\fnsep\thanks{\email{sofia@lip.pt}} on behalf of the Pierre Auger Collaboration\thanks{Authors list:\email{http://www.auger.org/archive/authors_2018_10.html}}\inst{2}\fnsep\thanks{\email{auger_spokespersons@fnal.gov}}
}

\institute{
LIP, Av. Prof. Gama Pinto, 2, Lisboa, Portugal  \and Observatorio Pierre Auger, Av. San Mart{\'\i}n Norte 304, 5613 Malarg\"ue, Argentina
}

\abstract{
The average profiles of cosmic ray shower development as a function of
atmospheric depth are measured for the first time with the Fluorescence
Detectors at the Pierre Auger Observatory. The profile shapes are well reproduced by the Gaisser-Hillas parametrization at the 1\% level in a 500 g/cm$^2$ interval around the shower maximum, for cosmic rays with 
$\log(E/\rm{eV})>17.8$. The results are quantified with two shape parameters, measured as a function of energy.

The average profiles carry information on the primary cosmic ray and its high energy hadronic interactions. The shape parameters predicted by the commonly used models are compatible with the measured ones within experimental uncertainties. Those uncertainties are dominated by systematics which, at present, prevent a detailed composition analysis.
}

\maketitle
\section{Introduction}

The Fluorescence Detector (FD) of the Pierre Auger Observatory~\cite{bib:auger} has collected unprecedented statistics of high quality data, imaging the longitudinal profile of the electromagnetic component of showers induced by ultra-high energy cosmic rays. The integral of the shower profiles gives direct calorimetric measurements of the primary particle energy, with small corrections due to the energy carried away by muons and neutrinos~\cite{bib:invisible}. The depth at which the profile maximum occurs, $\Xmax$, is the main
observable for analysis of the mass composition of primary cosmic rays. 

This work follows the hybrid data reconstruction procedure and event selection developed earlier for a composition unbiased $\Xmax$ measurement~\cite{bib:xmax}. The determination of each shower geometry uses the timing of the triggered FD pixels and one station in the Surface Detector, which should be at less than 1.5 km from the shower core, and have an expected trigger probability greater than 95\%, for proton and iron showers. The energy deposit profile is obtained from the photons detected at the telescope, taking into account the different emission mechanisms and the light absorption and scattering in the atmosphere; only events with viewing angles above 20$^\circ$ from the shower axis are accepted, to avoid direct Cherenkov light affecting the determination of $\Xmax$. In addition, a fiducial field of view is defined as a function of energy to ensure a uniform acceptance of most of the $\Xmax$ distribution observed in the data; finally, for each shower track, at least 300~g/cm$^2$ must be observed. This should include the $\Xmax$ position, for which the expected resolution, given the geometry, must be under 40~g/cm$^2$. 

The main goal of this contribution is to present a precise measurement of the shape of the longitudinal electromagnetic profile of cosmic ray showers~\cite{bib:paper}, and of the remaining information that it can carry about the interactions of primary cosmic rays.

\section{Measuring the profile shape}
\label{sec-1}

\begin{figure}
\hspace{+0.5cm}\includegraphics[width=0.475\textwidth]{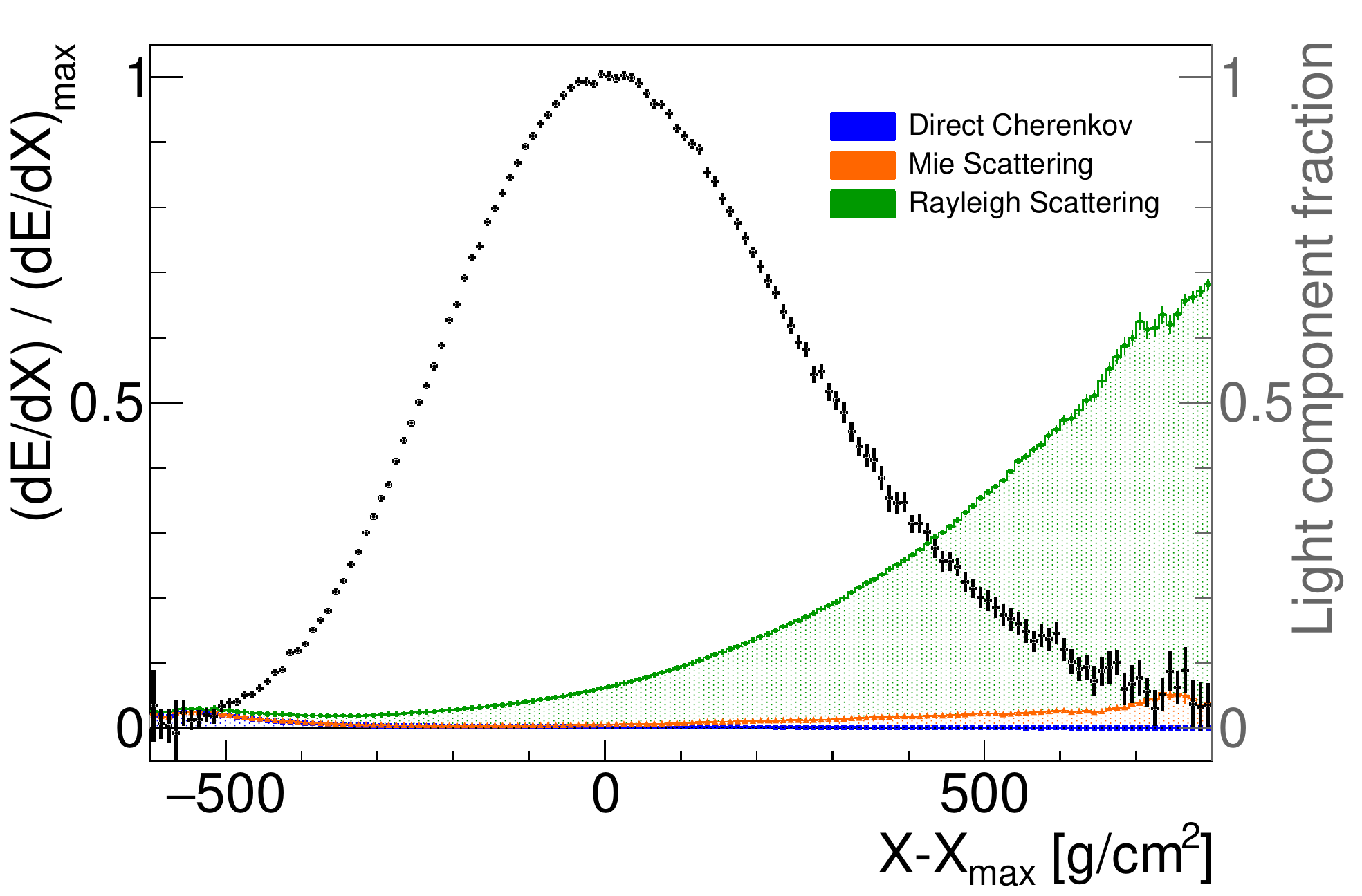}
\hfill
\hspace{-0.5cm}\includegraphics[width=0.475\textwidth]{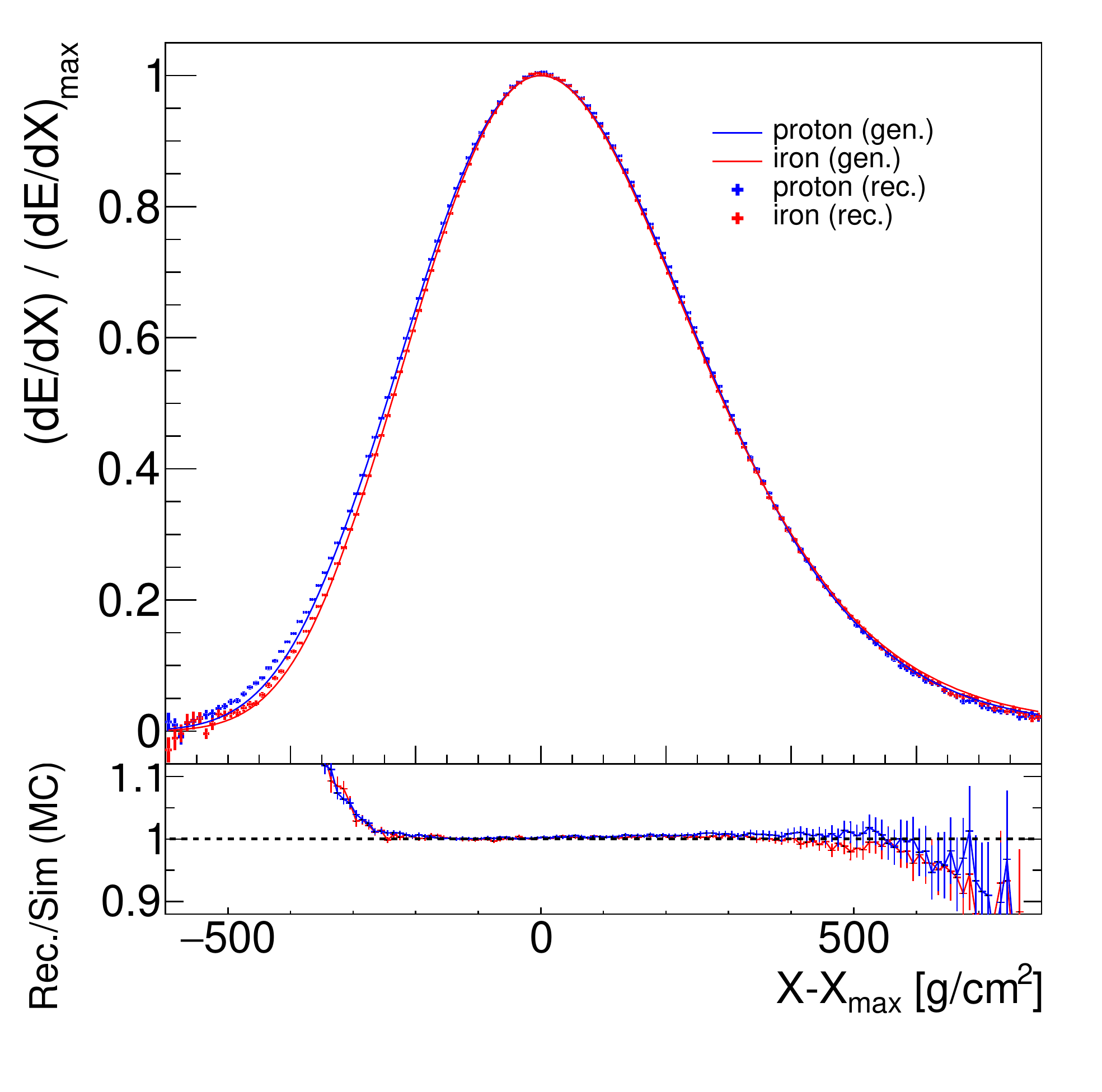}
\caption{\label{fig:aver_shower}Average longitudinal shower profile in data (top) and simulation (bottom) for events with energies between $10^{18.8}$ and $10^{19.2}$\,eV.
Top: The measured normalized energy deposit is shown in black, and the coloured regions (detailed in the legend) represent the average fraction of direct and scattered Cherenkov light in the photons measured at the telescope aperture, computed in the individual shower reconstruction. 
Bottom: Both generated and reconstructed profiles are shown (and their ratio in the inset plot) in blue for proton and red for iron simulated showers.}
\end{figure}

Because individual events are subjected to high statistical uncertainties, we construct a high accuracy average profile that can be analysed in detail and provide information about the sample. The profile of each event is normalized to its maximum energy deposit $\dEdXmax$, and centred at $\Xmax$. For each 10~g/cm$^2$ bin of $X'=X-\Xmax$, the normalized energy deposit is averaged, taking into account the respective uncertainties of each contributing event.

Figure~\ref{fig:aver_shower}(top) shows a reconstructed average profile in data, and the fractions of light components contributing at each $X'$. After the geometry of each event has been obtained, the light arriving at the FD at a given time is converted to its emission point at the shower axis, taking into account the attenuation by Rayleigh scattering in air and Mie scattering on aerosols. Most of the detected photons are from isotropic fluorescence emission, but a large component of the forward Cherenkov beam -- integrated along the shower axis -- can reach the telescope through scattering. 
The later part of the profile has a lower contribution of fluorescence light (the one directly proportional to the energy deposition) -- and so will be more subject to corrections of atmospheric effects and assumptions used in the reconstruction.

Figure~\ref{fig:aver_shower}(bottom) compares the average profiles of simulated energy deposition to the ones obtained after a full detector simulation and applying the same reconstruction procedure as used for data. The full chain is applied separately to proton and iron showers to show that the early part of the profile is the one that keeps information on the interaction of the different primary particles, and that this information is preserved by the reconstruction of the high accuracy average profile. 

\begin{figure}
\centering
\includegraphics[width=0.45\textwidth]{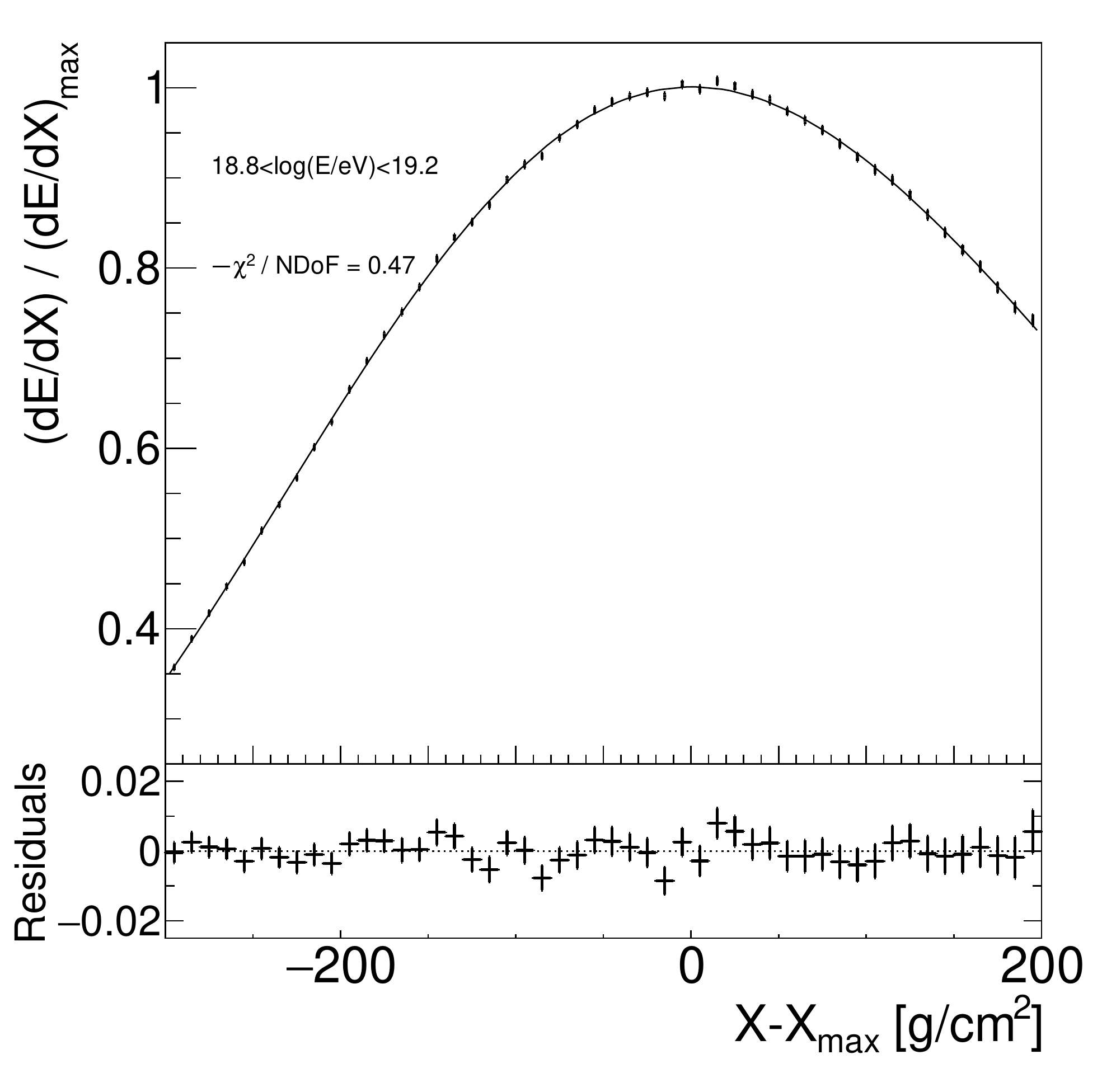}  
\caption{\label{fig:shortProf}Measured average longitudinal shower profiles for energies between $10^{18.8}$ and $10^{19.2}$\,eV. Data is shown together with the Gaisser-Hillas fit to the profile. The residuals of the fit are shown in the bottom inset.}
\end{figure}

Figure \ref{fig:shortProf} shows again the reconstructed average profile in data, in the region of $[-300,+200]$~g/cm$^2$ around the maximum, for which the detailed quantitative analysis is done. Also shown are the bin-by-bin residuals to the fitted Gaisser-Hillas function~\cite{bib:gh}, as an experimental test that it is a good representation of the shower longitudinal energy deposit profiles. The Gaisser-Hillas shape is confirmed at a 1\% level in each of six energy bins with $\log(E/\rm{eV})>17.8$. 

The Gaisser-Hillas function is written as~\cite{bib:lr}:
\begin{equation}
\label{usprl}
\left(\dEdX\right)' = \left(1+R \, \frac{X'}{L}\right)^{R^{-2}} \exp\left(-\frac{X'}{R \, L}\right)
\end{equation}
where $R=\sqrt{\lambda/|X'_0|}$, $L=\sqrt{|X'_0|\lambda}$ and $X'_0\equiv X_0 - \Xmax$. $L$ can be seen as a Gaussian width, and $R$ an asymmetry parameter,
with smaller correlations than the $\lambda$ and $X_0$ parameters that are more commonly used. 
These parameters keep information even when applied to average profiles~\cite{bib:ruben}.

\subsection{Systematic uncertainties}
\label{sec-2}

Table \ref{tab:syst} shows the maximum deviation caused by several categories of systematic uncertainties, compared to the much higher statistical accuracy. Most of the effects on the shape parameters are asymmetric and energy dependent. In the table only the highest values are presented.

The small differences found between the simulated and reconstructed profiles (fig.\ref{fig:aver_shower}) do not change significantly the values of $R$ and $L$ obtained in the chosen fit region, except for the first energy bin (up to 1~EeV). A small bias correction is determined by averaging the effect over proton and iron to return to the values as obtained with the generated $\dEdX$ values. Half of this correction is added as a systematic uncertainty, presented together with the effect of the energy scale uncertainty~\cite{bib:escale}.

The FD consists of 24 telescopes viewing elevations from 2 to 32 degrees, overlooking the 3000~km$^2$ Surface Detector (SD) which samples the particles arriving at ground with an array of stations separated by 1.5 km\footnote{Three extra telescopes observe higher elevations, and are complemented by extra SD stations with smaller separations - these are dedicated to lower energies not used in this work.}. One of the telescopes was excluded from the start of the analysis, since it showed large time residuals in the geometry fit. The stability of the average profile across the remaining 23 telescopes was used to check for detector effects, and the differences accounted for as a small detector systematic uncertainty. The effect of the hybrid geometry reconstruction is studied directly by varying within their uncertainties five independent parameters, representing the shower detector plane and the shower axis. Possible variations on the average shower profiles obtained with different selections in shower zenith angle or distance to the telescope were searched for, but found to be contained within the previously derived geometric systematic uncertainties in $L$ and $R$.

The determination of single event profiles, and their $\Xmax$ and $\dEdXmax$, is clearly a very fundamental piece of the analysis. In particular, the Gaisser-Hillas fit of single event profiles uses constraints to guarantee the convergence when short tracks are observed, affecting mainly low energy events. These were derived earlier from a low statistics sample of events with long tracks, and are expressed as $\lambda=L\cdot R = (61 \pm 13)$~g/cm$^2$ and $X_0=\Xmax+L/R=(121 \pm 172)$~g/cm$^2$; an additional constraint on the normalized integral was derived from simulations, compatible with all available models and compositions. Changing the constraints coherently by one standard deviation in each parameter results in systematic deviations of up to 3~g/cm$^2$ in $L$ and 0.01 in $R$. 

The fluorescence and Cherenkov yields were also varied within uncertainties and the reconstruction was repeated with and without multiple atmospheric scattering corrections, but a larger effect comes from separating events according to the percentage of light associated with direct fluorescence light: when selecting only events with more than 10\% of non-direct fluorescence light (the mean value in the analysed sample), the resulting average profile is wider, with $L$ increased by up to 2~g/cm$^2$ (while the change in $R$ is negligible).  

The atmospheric effects play an important role in the profile reconstruction and the Pierre Auger Observatory also has a large set of instruments dedicated to monitoring the atmospheric conditions~\cite{bib:auger}. The data was separated according to the seasons of the year, and the impact of cloud patches in the sky was also studied. The main effect comes, however, from the overall aerosol content and height profile. The atmospheric aerosol attenuation, $\tau_A(h)$, is obtained hourly by firing a laser from the centre of the SD array~\cite{bib:clf,bib:aero} and comparing the observed number of photons in the FD with the ones observed in a clear reference night. The correlated uncertainties in the definition of the clear night and the uncorrelated uncertainties from the hourly measurements were propagated by reconstructing again the full sample, and they lead to the largest systematic effect in the values of $L$, that changes by $\pm$ 5~g/cm$^2$, and $R$, that changes by $\pm$ 0.02.

\begin{table}[h]
{
\centering
\begin{tabular}{l@{\hskip 0.3in}c@{\hskip 0.3in}c}
                        & $\boldsymbol{R}$   & $\boldsymbol{L}$ [g/cm$^2$] \\
\hline
Atmosphere              & 0.030  & 5.5 \\ 
Light components \& fit & 0.018  & 3.3 \\ 
Geometry                & 0.018  & 2.2 \\
Detector                & 0.012  & 1.8 \\  
Bias corr. \& Energy scale    & 0.010  & 1.0 \\ 
\textbf{Total}         & \textbf{0.040}  & \textbf{7.3} \\
\hline
Statistical   & 0.012  & 0.9 \\
\hline
\end{tabular}
}
\caption{Breakdown of systematic uncertainties for $R$ and $L$. Uncertainties are energy dependent and asymmetric, so only the largest value is reported.}
\label{tab:syst}
\end{table}

\section{Two new observables}

\begin{figure*}         
\centering
\includegraphics[width=0.47\textwidth]{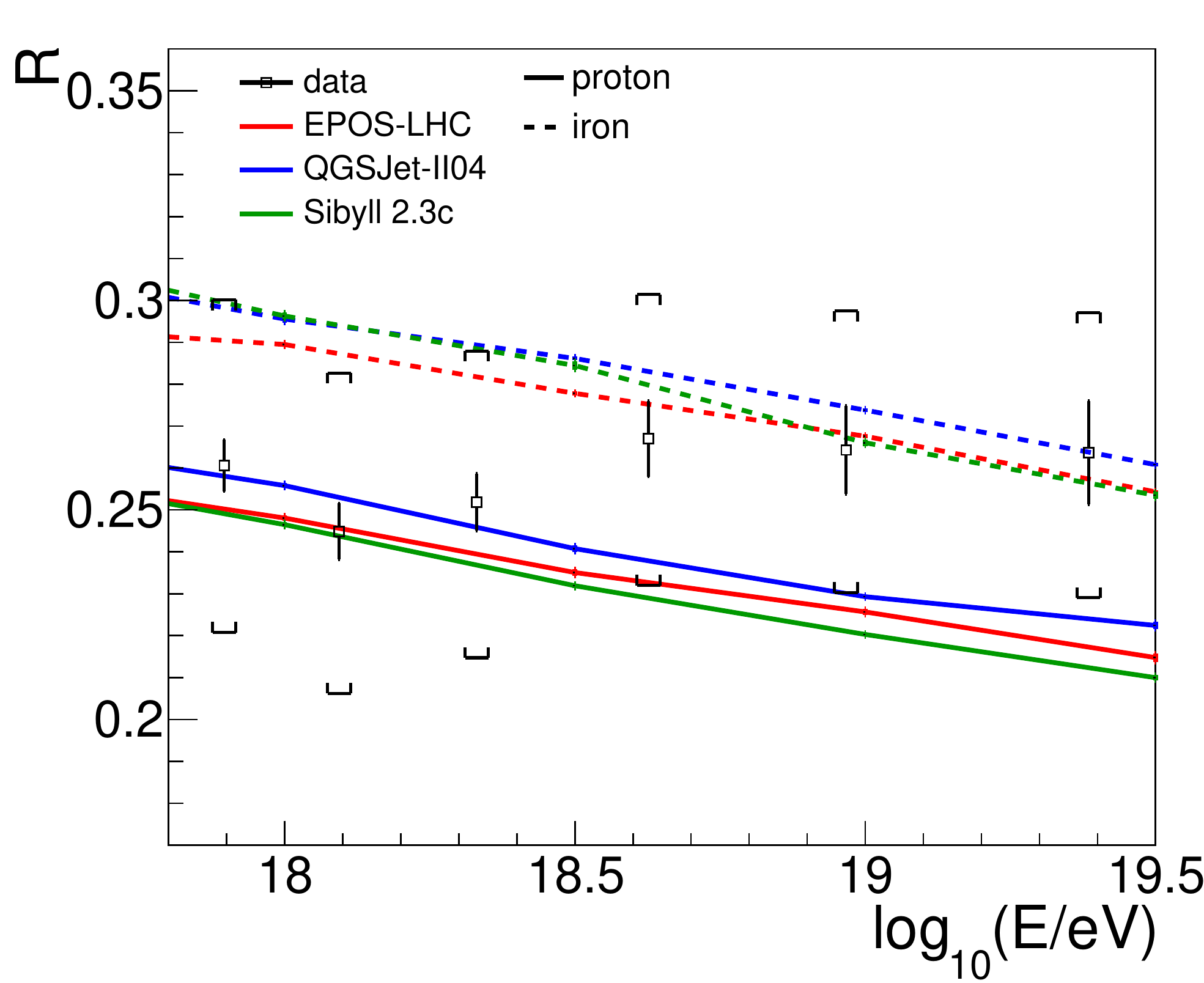}\hfill
\includegraphics[width=0.47\textwidth]{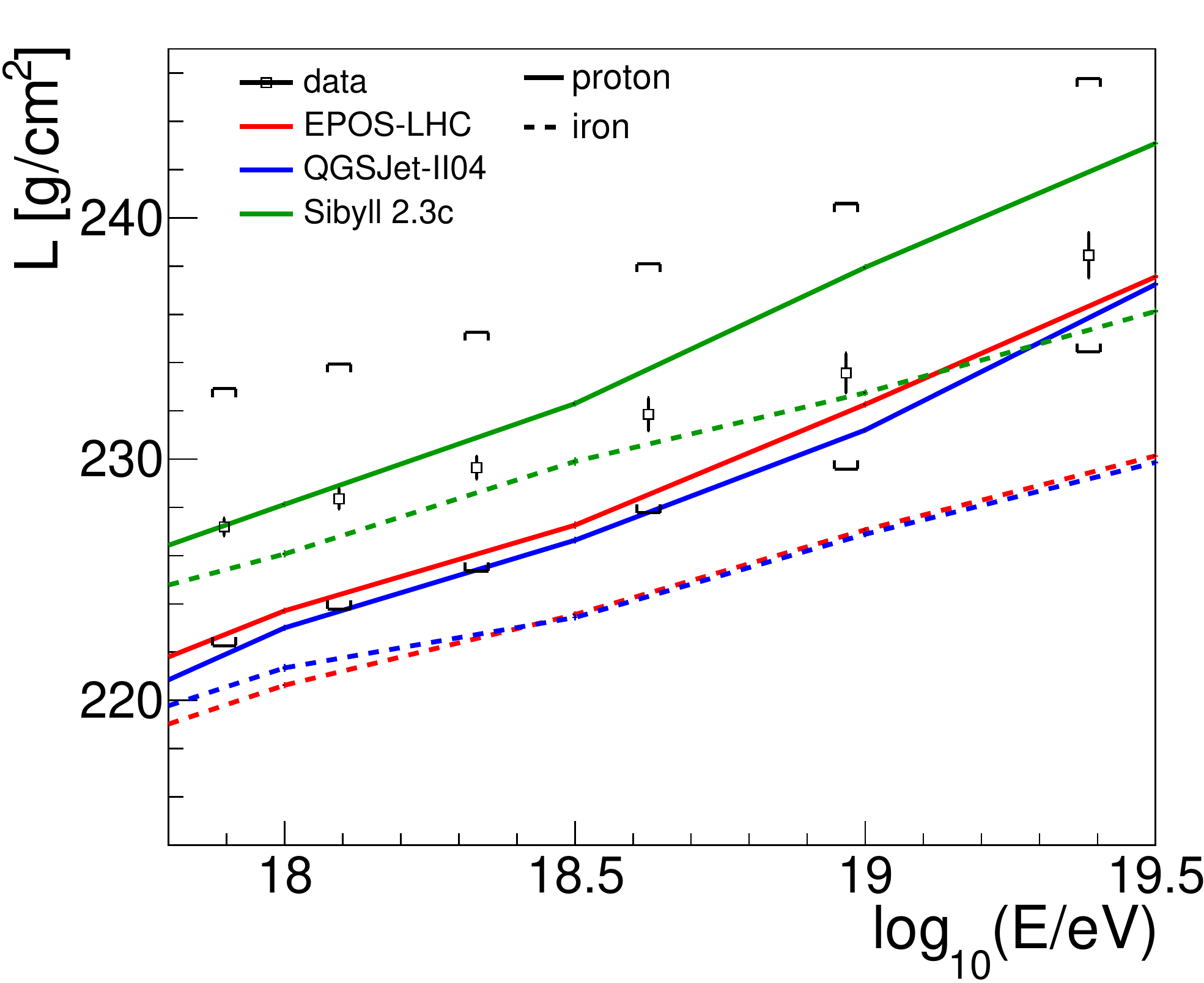}
\caption{$R$ (left) and $L$ (right) as a function of energy. The data are shown in black, with the vertical line representing the statistical error and the brackets the systematic uncertainty. Air showers were simulated with CORSIKA~\cite{bib:corsika} using different interaction models (see legend). The predictions from simulations are shown with full lines for proton and dashed lines for iron primaries.}
\label{fig:canvas_rl}
\end{figure*}

Figure \ref{fig:canvas_rl} shows the measured values of $L$ and $R$, compared to the predictions of three hadronic interaction models for proton and for iron primaries. We conclude that the measured shower shapes are compatible with model predictions within uncertainties and so that the models provide a fair description of the electromagnetic shower component, measurable by the Fluorescence Detectors.

The predictions of the models for the $L$ variable are different, with Sibyll~2.3c~\cite{bib:sibyll} predicting higher values than either QGSJetII-04~\cite{bib:qgsjet} or EPOS-LHC~\cite{bib:epos}. The data is compatible with both proton or iron values in Sibyll~2.3c, while it is above the prediction for pure iron from the other two models. On the other hand, the three models predict similar values for $R$, higher for iron than for proton showers, and decreasing with energy for each pure composition. In this case, the data indicates a slower energy evolution, but the present systematic uncertainty prevents an analysis in terms of primary mass composition. 

\begin{figure*}         
\centering
\includegraphics[width=0.49\textwidth]{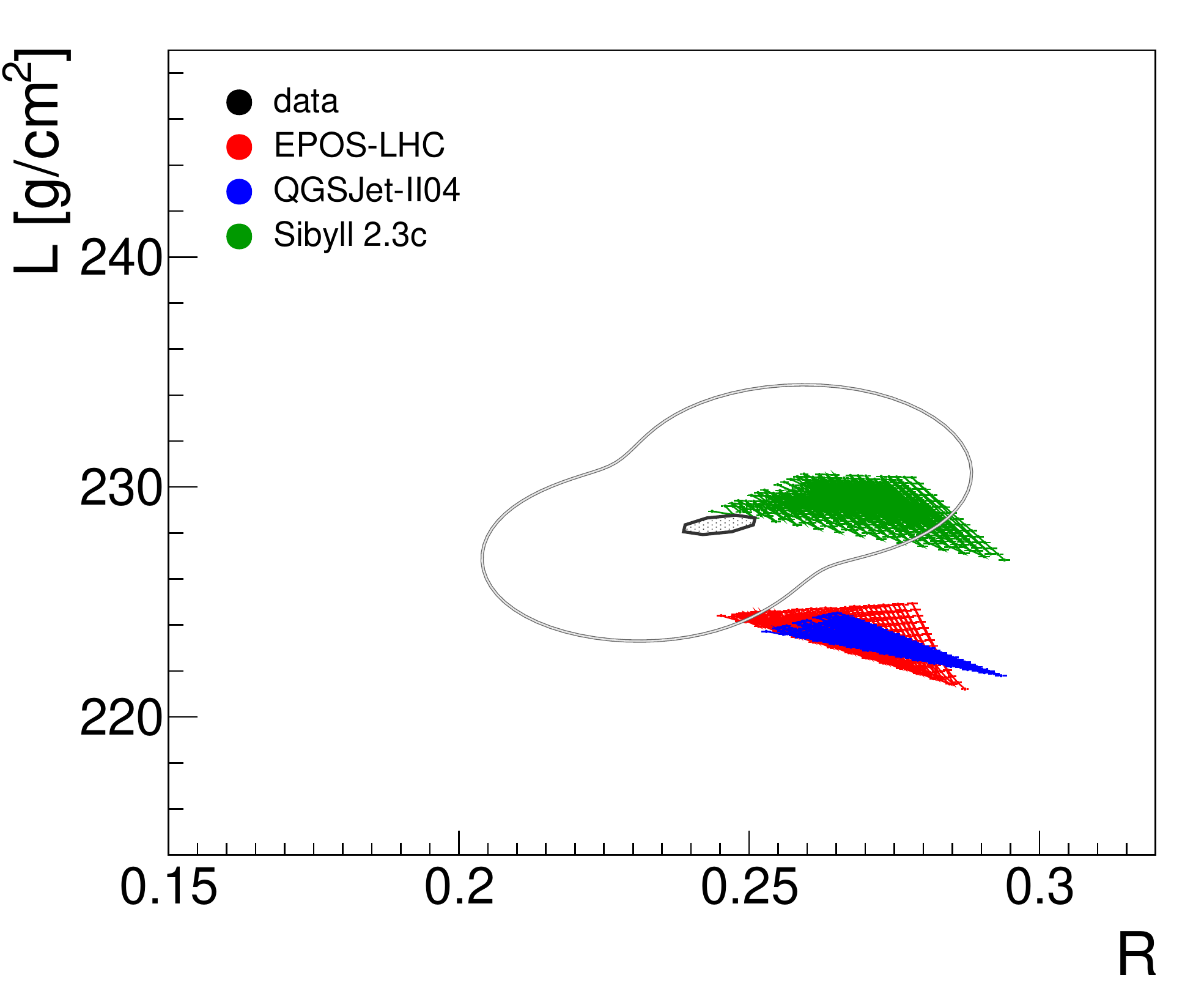}\hfil
\includegraphics[width=0.49\textwidth]{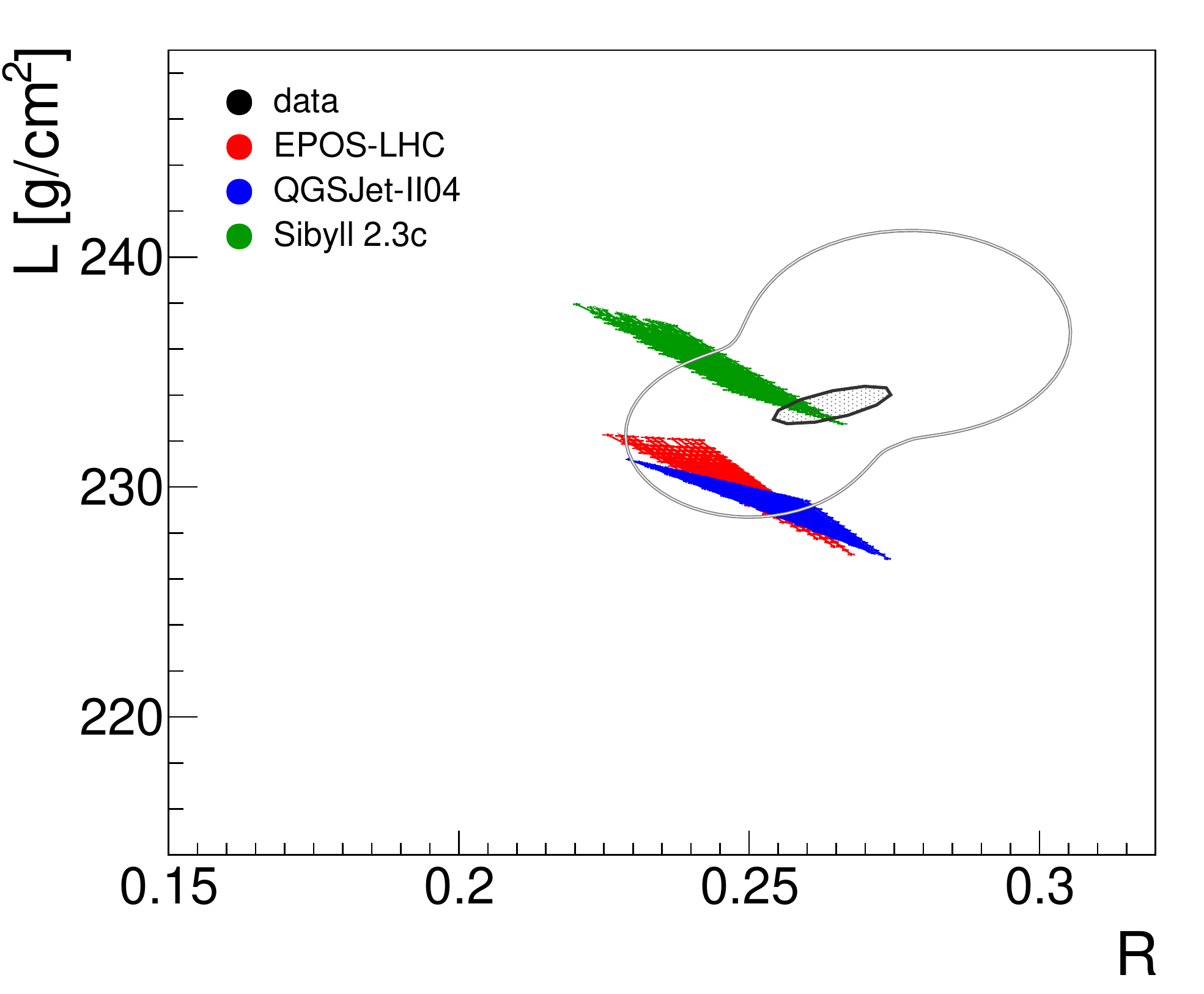}    
\caption{$L$ vs $R$ for the energy bin $10^{18}$ to $10^{18.2}$\,eV (left) and from $10^{18.8}$ to $10^{19.2}$\,eV (right). The inner dark grey ellipse shows the fitted value for data and its statistical uncertainty, and the outer light grey area the systematic uncertainty. For each hadronic model proton, helium, nitrogen and iron showers were simulated and average profiles were built making all possible combinations. Each of the points represents the value of $L$ and $R$ for a given model and composition combination, so the phase space spanned by each model is contained in its respective coloured area. Pure proton is, for each model, on the upper left side (high $L$ and low $R$) and the transition to iron goes gradually to the lower right side.}
\label{fig:canvas_2d}
\end{figure*}

Figure \ref{fig:canvas_2d} puts together the two observables in two energy bins, highlighting the correlations of statistical and systematic uncertainties. It also shows the prediction for all possible mass compositions (obtained by combining different fractions of H, He, Ni and Fe), according to each of the high energy hadronic interaction models. The combination of the two variables allows a clearer separation for different predictions. For example, in the energy bin of $\log{(E/\rm{eV})}\in[18,18.2]$, the profile shape parameters predicted by QGSJet-II04 for any composition are at more than one standard deviation from the values allowed by data, while the shape predicted by Sibyll~2.3c is fully compatible with the measurements for both energies. A measurement of the average shape with lower systematic uncertainties would provide constraints on the hadronic models, specially when put together with other composition sensitive variables.

\section{Summary}

We present the first measurement of the shape of the average shower profiles as a function of atmospheric depth, with high statistical precision, and a 1\% level check that these follow the Gaisser-Hillas parametrization in the region of [-300, 200]~g/cm$^2$ around the shower maximum.

The analysis provides a systematic check of the shower reconstruction procedures used in the Pierre Auger Observatory, and it is concluded that the present measurement of the profile shape parameters is mostly affected by the uncertainties in the description of the height profiles of atmospheric aerosols.

The predictions of the hadronic interaction models for the longitudinal profile of the electromagnetic shower component are compatible with the measurements, within present experimental uncertainties. However, it is also shown that the two new observables ($L$ and $R$) that describe the average profile shape have the potential of further discriminating between models and mass composition scenarios.

%
\section{Acknowledgements}
The presenter acknowledges the grants CERN/FIS-PAR/0023/2017 and IF/01724/2013.

\end{document}